\documentclass[prl,aps,twocolumn,showpacs]{revtex4}
\usepackage{bm}
\usepackage{epsfig}

\usepackage{amsmath}
\usepackage{amssymb}
\usepackage{color}

\newcommand{\ee}{\mathrm{e}}

\newcommand{\CC}{\mathcal{C}}

\begin{document}

\title{Investigating amorphous order in stable glasses by random pinning}
\author{Christopher J. Fullerton}
\author{Robert L. Jack}
\affiliation{Department of Physics, University of Bath, Bath, BA2 7AY, United Kingdom}

\begin{abstract}
We use a random pinning procedure to investigate stable glassy states associated with
large deviations of the activity in a model glass-former.  We pin particles both from
active (equilibrium) configurations and from stable (inactive) glassy states. By comparing the distribution of the overlap
between states that share pinned particles, we infer that the
inactive states are characterised by a structural length scale that is comparable
with the system size.  This is a manifestation of amorphous order in
these glassy states, which helps to explain their stability.
\end{abstract}


\maketitle

Glassy materials are stable and appear to be solid, but their molecular structures closely resemble those
of liquids~\cite{nagel-review,deb-review}.  Reconciling these two observations is a central challenge if the properties of these
important materials are to be understood.  To this end, a useful concept is
\emph{amorphous order}, which means that while the structure of a glass appears highly disordered,
there may nevertheless be strong correlations between particles, extending over
significant length scales~\cite{KL,BB-ktw-2004,montanari06}, and leading to glassy behavior~\cite{adam-gibbs,ktw,BB-ktw-2004}.  
These correlations
can be revealed (for example) by point-to-set correlation functions~\cite{berthier-pin2012,biroli-pin,berthier-kob-pin2013,
cavagna08,berthier-silvio}.
Alternatively,
given that the most significant differences between liquid and glass states 
appear in \emph{dynamical} measurements, one may argue
that a useful description of the glass
transition should focus on particle dynamics~\cite{GC-2003,GC-annrev-2010}.
Recent studies based on this hypothesis have
revealed non-equilibrium phase transitions~\cite{kcm-transition,hedges09} that occur when systems are biased to suppress particle motion,
in ``$s$-ensembles''. 
Here, we use random pinning (point-to-set) 
measurements~\cite{berthier-pin2012,biroli-pin,berthier-kob-pin2013} to
show that while these stable glasses were found by analysing their dynamical properties, they
also exhibit strong amorphous order.  
Hence, we argue that the dynamical approach of
the $s$-ensemble and the structural idea of growing amorphous order are not contradictory~\cite{jack10-rom}:
rather, they offer complementary routes~\cite{biroli-garrahan} 
by which theories of the glass transition may be developed and refined.

We consider the well-studied glass-forming liquid 
of Kob and Andersen~\cite{ka95}.  
As well as its equilibrium state, this system exhibits a non-equilibrium `inactive phase'~\cite{hedges09}, which is
extremely stable~\cite{jack11-stable}, and is found by biasing dynamical trajectories to lower than average activity.
The stable glassy states that we consider were taken from this inactive phase,
for a system of $N=150$ particles.  Full details are given in Supplementary Information (SI).    
The natural unit of length in the system is the diameter $\sigma$ of the larger particles (species A), the system
size is $L=5\sigma$, and  
all results shown are for temperature $T=0.6$ (in units of the AA-interaction energy), for which the equilibrium
state is a weakly supercooled fluid.  The system evolves
by overdamped (Monte Carlo) dynamics~\cite{berthier-mc2007}, which
gives results for structural relaxation
in quantitative agreement with molecular dynamics~\cite{berthier-mc2007,hedges09}.
Time is measured in units of  $\Delta t=\sigma^2/D_0$, where
$D_0$ is the diffusion constant of a free particle.

To analyse the connection between amorphous order and the stability of inactive states, we
use a random pinning procedure~\cite{berthier-pin2012,biroli-pin}. For a given \emph{reference configuration},
we fix the position of each particle with probability $c$, arriving at a \emph{template}: a set of approximately $cN$ 
pinned particles.  The remaining (unpinned) particles then move as normal in the presence of the frozen template.
If the reference configuration comes from a highly-ordered state, 
one expects a strong influence of the template on the dynamic and thermodynamic properties of the system.
For example, in a perfectly crystalline sample, a template containing just 
three particles is sufficient to determine the lattice orientation and hence the positions of all other particles.
More generally, if a template containing a small fraction of particles has a strong influence on the liquid structure,
this indicates that the correlations among particle positions are strong, 
and hence that the system is ordered, even if this order is not apparent from (for example) two-point density correlations.

To analyse the influence of the template, we require a measure of similarity between configurations.  
To obtain useful measurements for configurations where particle indices are permuted but the structure
remains similar, we divide the system into a cubic grid of cells of linear
size $\ell=(L/10)=0.5\sigma$~\cite{berthier-pin2012}. 
Let $n_i$ be the number of mobile (unpinned) particles of type A in cell $i$.  Then, if configurations $\CC$ and $\CC'$ have
cell occupancies $\{n_i\}$ and $\{n_i'\}$, their
(normalised) overlap is
\begin{equation}
Q(\CC,\CC') = \frac1M \sum_i \frac{ n_i n_i' - \langle n_i \rangle^2 }{ \langle n_i^2\rangle - \langle n_i \rangle^2} .
\end{equation}
If $\CC$ and $\CC'$ are identical then $\langle Q\rangle = 1$ while
for independent random configurations $\langle Q \rangle = 0$.

In equilibrium studies of pinning~\cite{berthier-pin2012,biroli-pin,berthier-kob-pin2013,charb-prl2012,krakoviack2010,kurzidim2011,kim-pin2011,karmarkar2012,jack-plaq-pin,jack-pin-chi4} one begins by
drawing an equilibrium reference configuration $\CC_0$, from which 
each particle is pinned with probability $c$.  Then, a second configuration $\CC$ is generated, which
includes the pinned particles from $\CC_0$, while the remaining particles are equilibrated
in the presence of this template. 
On repeating this procedure many times, one may build up a distribution
$p_{\rm eq}(Q|c)$ for the overlap $Q=Q(\CC_0,\CC)$.   The associated ensemble is discussed in SI.
We emphasise that this is a static (thermodynamic)
procedure, in that $p_{\rm eq}(Q|c)$ depends 
only on the Boltzmann distribution (or potential energy surface) of the system.

To measure amorphous order for inactive states,
we draw reference configurations $\CC_0$ from 
trajectories of the model that are biased
to lower than average dynamical activity -- full details are given in SI.
The inactive state
is in contact with a thermostat at $T=0.6$ at all times, so the natural comparison is between  the
inactive state and an equilibrium state at that temperature.
Starting from the inactive reference configurations, we use the same pinning procedure as described above, 
which results in a different distribution of the overlap, denoted by
$p_{\rm in}(Q|c)$.  If the inactive states have increased amorphous order as compared to equilibrium, one expects
significant differences between $p_{\rm eq}(Q|c)$ and $p_{\rm in}(Q|c)$.

\begin{figure}
\includegraphics[height=4cm]{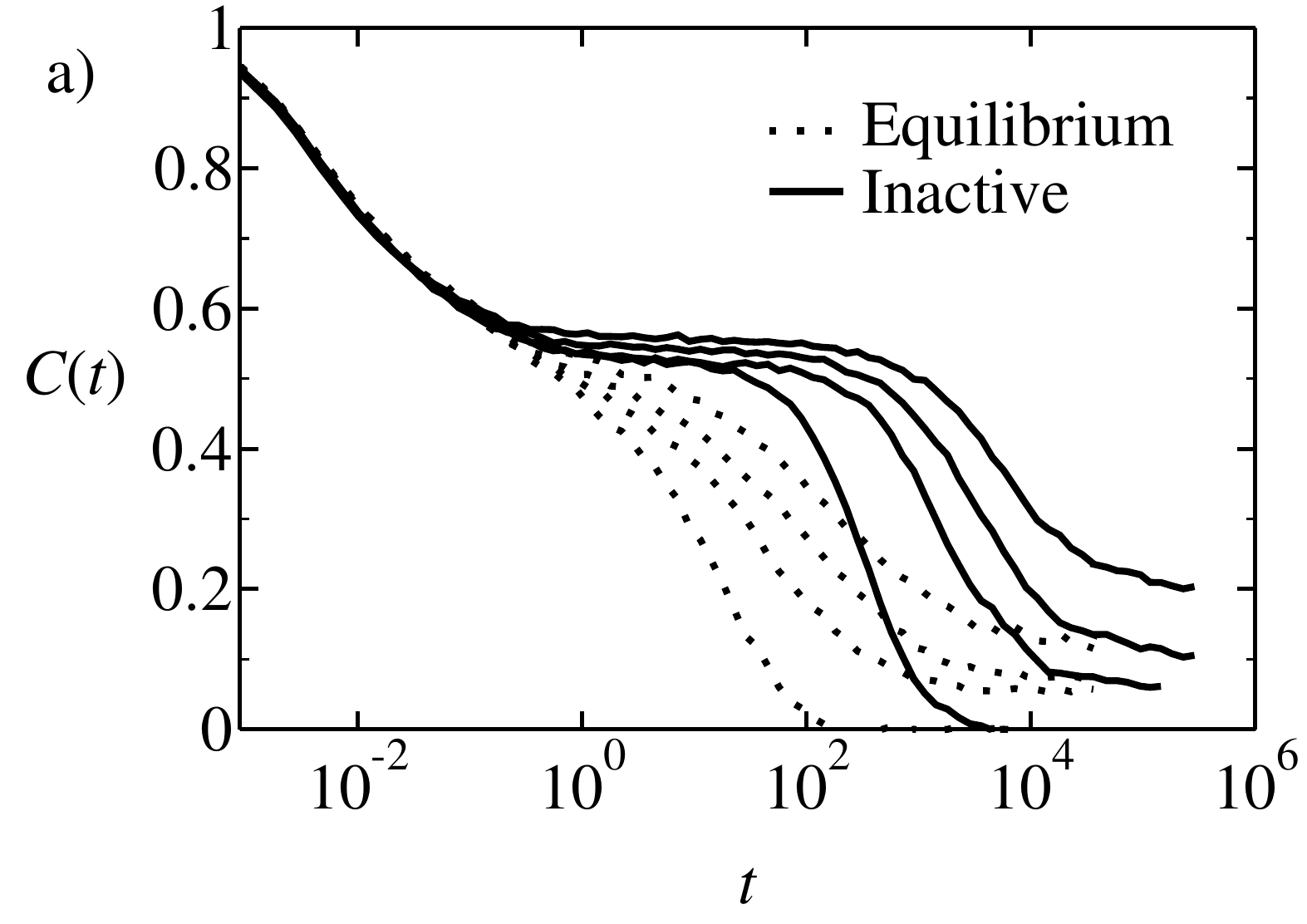}\par
\vspace{4pt}
\includegraphics[height=4.2cm]{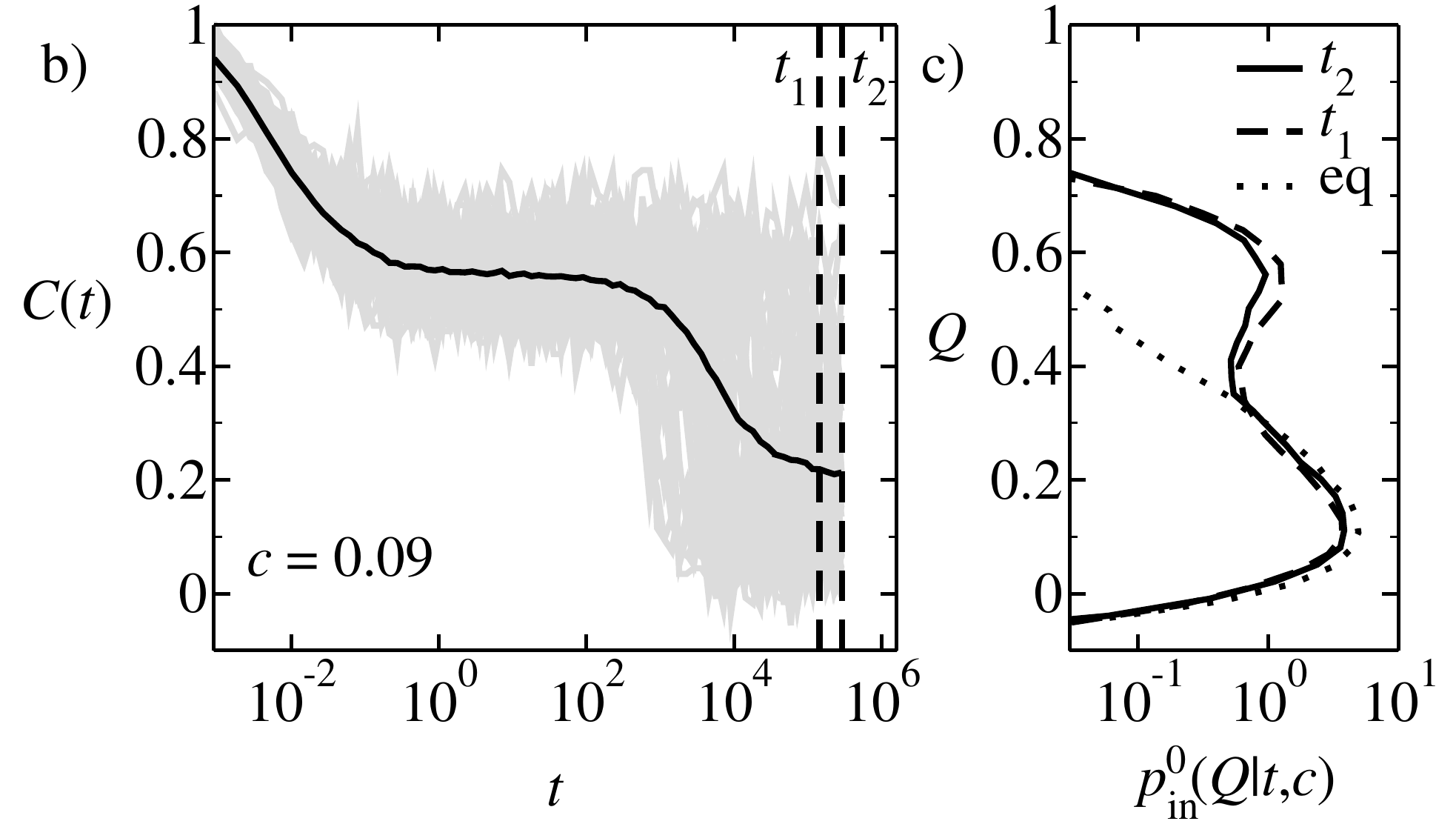}\par
\vspace{4pt}
\includegraphics[height=4.2cm]{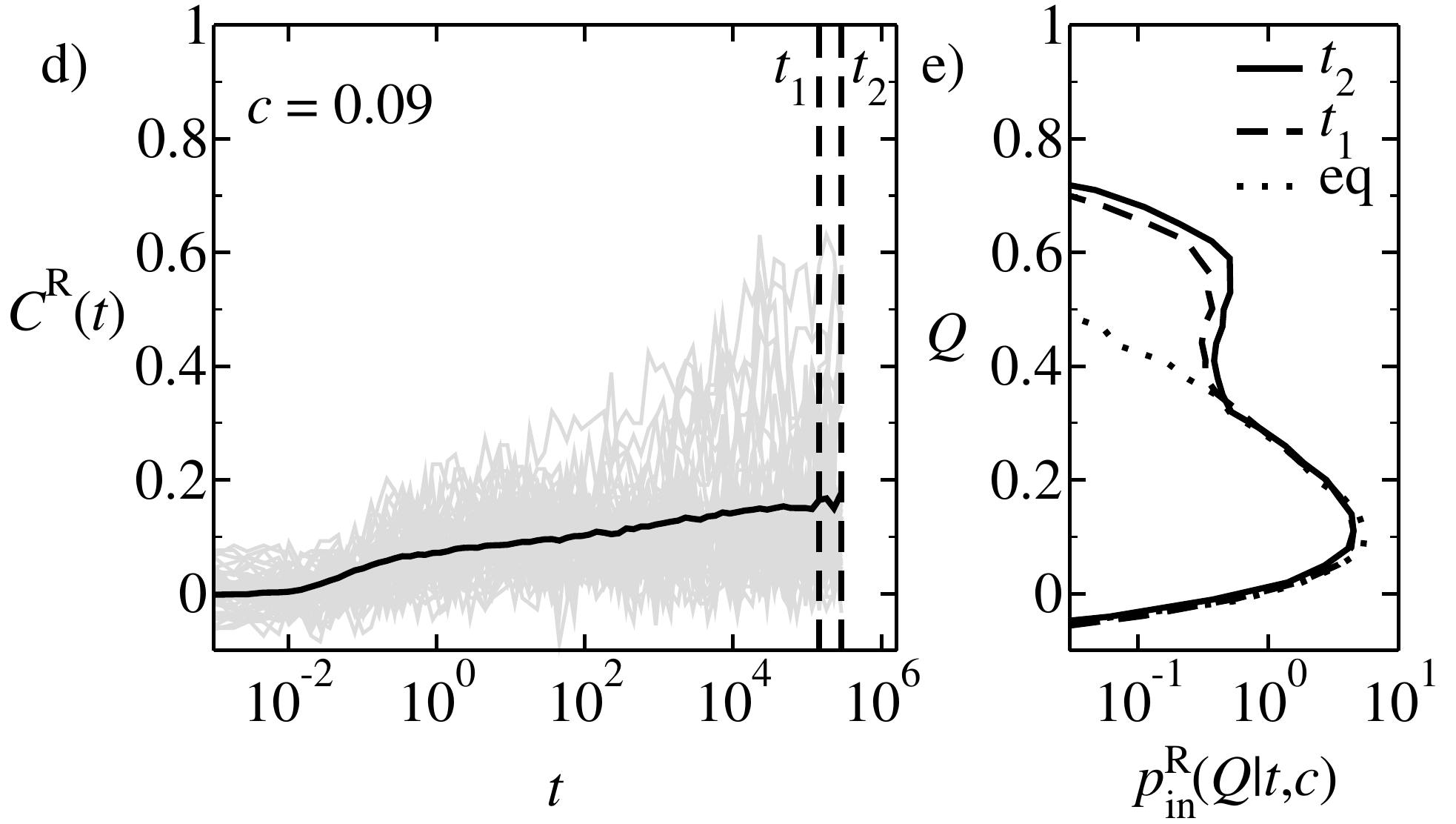}
\caption{({\bf a})~$C(t)$, calculated from equilibrium and inactive reference configurations. From left to right $c = 0.00, 0.05, 0.07, 0.09$.
({\bf b})~For an inactive reference state and $c=0.09$, the average overlap $C(t)$ (black) 
and plots of $Q(\CC_0,\CC_t)$ for representative trajectories (gray). 
({\bf c})~Distributions of $Q(\CC_0,\CC_t)$ (gray data from (b)) at the times indicated, compared with $p^0_{\rm {eq}}(Q|t,c)$
for an equilibrium reference state, at $c=0.09$ and $t=t_2/4$, representative of the long-time limit.  
({\bf d},{\bf e})~Data analogous to (c,d), except that 
dynamical trajectories were started from configurations in which the positions of unpinned particles were `randomised' at $t=0$
(see text).
For all histograms, results are accumulated over time windows ending at the indicated times: the windows are long enough to enable precise
measurements but short enough that they do not significantly affect the shapes of the distributions.
}
\label{fig:Qt}
\label{fig:vap}
\end{figure}

To infer the form of $p_{\rm in}(Q|c)$ and $p_{\rm eq}(Q|c)$, we conducted dynamical simulations.  
For a given reference configuration $\CC_0$
and a given template, we simulated dynamical trajectories starting from $\CC_0$. This leads to a time-dependent
overlap $Q(\CC_0,\CC_t)$, where $\CC_t$ is the configuration of the system at time $t$.
We then repeat the procedure for many different templates and different reference configurations.
Fig.~\ref{fig:Qt}(a) shows the time-dependent average overlap $C(t) = \langle Q(\CC_0,\CC_t) \rangle$, 
comparing the behavior for equilibrium and inactive reference configurations. 
As in~\cite{jack11-stable}, the dynamical relaxation from inactive states is much slower than equilibrium
relaxation, even in the absence of pinning.
Also, as $c$ is increased, the dynamical relaxation slows down, for both equilibrium and inactive reference 
states~\cite{jack-pin-chi4}.

Fig.~\ref{fig:Qt}(b,c) indicates that the slow decay of $C(t)$ is associated with large fluctuations of $Q(\CC_0,\CC_t)$.
For long times, the time-dependent distribution $p_{\rm in}^0(Q|t,c)$ of this overlap has a characteristic
bimodal shape.  In contrast, the distribution $p_{\rm eq}^0(Q|t,c)$, obtained under the same conditions,
lacks the second peak at high-$Q$.  The differences between these distributions are entirely due to the 
to the structural differences between the 
reference states (inactive or equilibrium) from which the pinned particles were selected.  Further, the dynamics used here ensure (see SI) that
$\lim_{t\to\infty}p_{\rm in}^0(Q|t,c)= p_{\rm in}(Q|c)$, which indicates that differences between the long-time
limits of $p_{\rm in}^0(Q|t,c)$ and $p_{\rm eq}^0(Q|t,c)$ can be attributed to amorphous order, as measured by
$p_{\rm eq}(Q|c)$ and $p_{\rm in}(Q|c)$.  However,
 it is clear from Fig.~\ref{fig:Qt} that $p_{\rm in}^0(Q|t,c)$ has not reached
its large-$t$ limit, so we may not assume that the 
`dynamical' distribution $p_{\rm in}^0(Q|t,c)$ reflects the form of the `static'
distribution of interest, $p_{\rm in}(Q|c)$.  
In particular, the secondary maximum at large-$Q$ in $p_{\rm in}^0(Q|t,c)$ might disappear on increasing $t$, as systems
finally relax away from the reference configuration into other available states. 

To address this point, we conducted simulations in which a template was fixed as before,
after which the temperature was increased to $T = 5.0$ and dynamics run for $t \approx 1000\Delta t$.
This temperature is high enough that the mobile particles quickly decorrelate from their initial configuration.
These `randomised' states were then used as initial conditions for dynamical simulations (see~\cite{hocky2012}
for a similar idea).
As before, we measure the distribution of the overlap between the reference $\CC_0$ and the resulting time-dependent
configurations $\CC_t$.
Let the distribution of this overlap be $p_{\rm in}^{\rm R}(Q|t,c)$ and let $C^{\rm R}(t) = \langle Q(\CC_0,\CC_t) \rangle$ be
the average overlap for this distribution.  

Results are shown in Fig.~\ref{fig:vap}(d,e): the average overlap $C^{\rm R}(t)$ starts near zero (as expected for a randomised
initial condition) and slowly increases, due to the influence of the template.
Further, for large times, there are a substantial number of trajectories where the system spontaneously evolves into a state with large $Q$. 
This means that the frozen template (containing just $9\%$ of the particles)
influences the system strongly enough that it has a significant probability of returning to the 
metastable state associated with the original reference configuration.
As before, $p_{\rm in}^{\rm R}(Q|t,c)\to p_{\rm in}(Q|c)$ as $t\to\infty$ but this
limit is not saturated.  However, we see that while $p_{\rm in}^{\rm R}(Q|t,c)$ and $p_{\rm inac}^0(Q|t,c)$ are converging
to the same limit, they do so from opposite directions, in that the original simulations start
in the reference state $\CC_0$ and evolve away from it, while the `randomised' simulations start far from $\CC_0$
and evolve back towards this reference state.  Thus, a natural conjecture is that these
two distributions give (approximate) upper and lower bounds
on the limit distribution $p_{\rm in}(Q|c)$. 

\begin{figure}
\vspace{12pt}
\includegraphics[width=8.5cm]{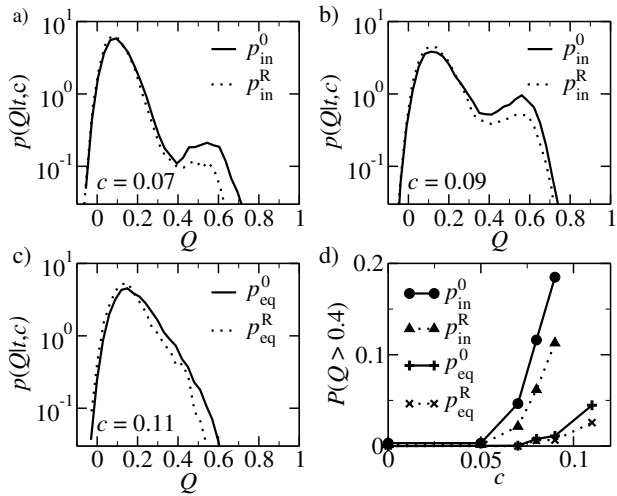}
\caption{({\bf a}-{\bf c})~Distributions of $Q(\CC_0,\CC_t)$. ({a})~$c = 0.07$ for an inactive reference; 
({b})~$c = 0.09$, inactive reference; ({c})~equilibrium reference, for the relatively high pinning probability $c = 0.11$
(results for lower $c$ are similarly unimodal).
In~(b), $t=t_2$ as indicated in Fig.~\ref{fig:Qt}b; 
for~(a,c) the times are $3t_2/4$ and $t_2/4$ respectively.  
({\bf d})~Probability that the overlap $Q > 0.4$, corresponding to the high-$Q$ peak in the distributions $p(Q|t,c)$, 
for equilibrium and inactive references.}
\label{fig:histo}
\end{figure}

Fig.~\ref{fig:histo} collates the relevant distributions.
The key observation
is that the inactive reference configurations lead to bimodal distributions of the overlap, while equilibrium reference configurations result in
unimodal distributions, even for pinning fractions as high as $c=0.11$.
We note also that the probability associated with the large-$Q$
peak in $p_{\rm in}(Q|t,c)$ rises in a strongly non-linear fashion, indicating the central
role of many-body correlations~\cite{szamel-pin2013}.


We now discuss how these numerical results shed light on the nature of  
amorphous order in these systems.
It is natural to write 
\begin{equation}
p(Q|c) = \ee^{-N\beta V(Q,c)}
\end{equation}
where $V(Q,c)$ is an effective potential, as used in mean-field theories of the glass transition~\cite{franz-parisi-1997}, 
generalised to include the effect of the frozen templates.  
Within mean-field theories and below the onset temperature
($T_{\rm o} \approx 1$ for this model) one
expects two peaks in $p(Q|c)$, as $Q$ is varied, and hence two minima in $V(Q,c)$. 
These correspond to the cases where $\CC$ and $\CC_0$
are in the same metastable state (high-$Q$) or in different states (low-$Q$).  As $c$ is increased, one expects the low-$Q$
peak to be reduced, because cases where $\CC$ is in a different state from $\CC_0$ are not typically consistent with the
frozen template.  
Within random first-order transition theory~\cite{ktw}, one additionally expects a phase transition at some critical 
concentration $c$~\cite{biroli-pin,camma-pin-alpha}, 
so that the high-$Q$ peak of $p(Q|c)$ dominates the distribution for $c>c^*$, while the low-$Q$ peak
dominates for $c<c^*$.   

If such phase transitions occur in randomly pinned systems, the distribution $P(Q|c)$ remains bimodal
as the system size $N\to\infty$.  Numerically, this can be inferred by a finite-size scaling analysis~\cite{berthier-kob-pin2013}.  
However, the long time scales associated with inactive states means that we have not been able to conduct such an analysis in this system.
Nevertheless, the bimodal
distributions $P(Q|c)$ and the associated non-convex $V(Q,c)$ shown here imply the existence of strong many-body correlations in these
systems.  As we now explain, we are able to infer from Fig.~\ref{fig:histo}
that the inactive state in this model has a structural
correlation length $\xi$ that is comparable with the system size $L=5\sigma$.  

\begin{figure}
\includegraphics[width=8.5cm]{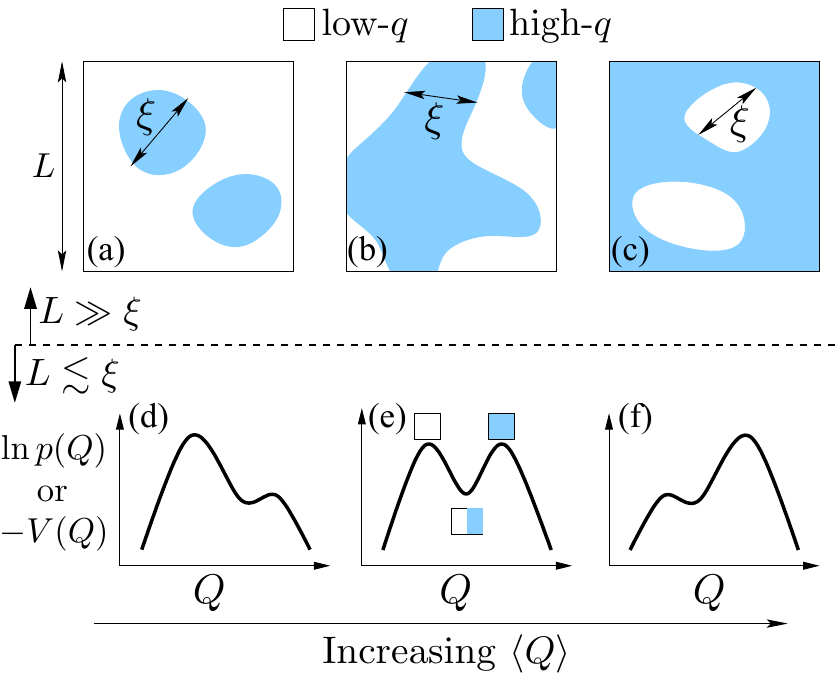}
\caption{({\bf a-c}) 
Sketches of the spatial distribution of the overlap $q(\bm{r})$ within a large system ($L\gg \xi)$, as the average value of
$Q$ increases.  One expects
a domain structure of high-$q$ and low-$q$ patches, with an associated length scale $\xi$, as indicated.  ({\bf d-f}) Sketches
of distributions $\ln p(Q) = - N\beta V(Q)$ for small systems ($L\lesssim \xi$), under the same conditions as (a-c).
For small systems then the domain structure
leads to two peaks in $p(Q)$, with typical realisations of the system containing only one domain, as indicated in (e).}
\label{fig:dom}
\end{figure}

To show this, 
we consider spatial fluctuations.
For an equilibrium reference state at the temperature considered
here, it is expected
that spatial fluctuations of the overlap prevent any phase transition~\cite{biroli-pin,jack-pin-chi4}.  
Hence, within the general framework of the renormalisation group,  one does not expect any long-ranged order
in the system, but one does expect strong spatial fluctuations of
the order parameter $Q$, with a finite correlation length~$\xi$.  
The expected situation is sketched in Fig.~\ref{fig:dom}.  This picture is realised (for example) in plaquette spin models~\cite{jack-plaq-pin},
which have glassy dynamics and growing amorphous order at low temperatures~\cite{jack-caging,camma-patch}.
Consider two configurations $\CC$ and $\CC'$ that share a template: 
shaded regions in Fig.~\ref{fig:dom}(a-c) indicate parts of the system where the overlap between $\CC$ and $\CC'$ is large.
Specifically, we
define a local overlap $q(\bm{r},\CC,\CC')$ 
so that $Q(\CC,\CC') \propto \int\! \mathrm{d}\bm{r}\, q(\bm{r},\CC,\CC')$.  The two-point
correlations of $q$ are characterised by
 $G_4(\bm{r}-\bm{r'}) = \langle q(\bm{r}) q(\bm{r}') \rangle - \langle q\rangle^2$, which is related~\cite{jack-plaq-pin} to the four-point
 correlation functions that have been studied extensively in glassy systems~\cite{DH-book}.  (Here $q$ is an overlap between
 two configurations, so the two-point correlations of $q$ correspond to four-point correlations of the underlying density field.)
We define $\xi$ as the length scale associated with the large-$|\bm{r}|$ decay of $G_4(\bm{r})$.

Figs.~\ref{fig:dom}(a-c) illustrate three cases where the correlation
length $\xi$ is non-trivial, over a range of $\langle Q\rangle$. 
In Fig.~\ref{fig:dom}(c),
$\langle Q \rangle$ is relatively large (strong pinning), and
most of the system has high-$q$, while
small-$q$ domains represent regions where the system has performed a localised relaxation process, and differs from  
the reference state.
As one decreases $\langle Q\rangle$ by reducing pinning [Fig.~\ref{fig:dom}(b)], more of the system is covered by
small-$q$ domains, with a characteristic
length scale $\xi$ (the situation is similar to the paramagnetic state of an Ising-like model).  On further reducing $\langle Q\rangle$,
[Fig.~\ref{fig:dom}(a)] the small-$q$ regions predominate, leaving behind high-$q$ domains where configurations $\CC$ and $\CC'$ are
similar, perhaps due to a random fluctuation, or to a particular property of the template in that area.
We note that while Figs.~\ref{fig:dom}(a-c) represent systems over a range of $c$, 
they are all quite far from the limiting cases of strong pinning ($c\to1$), where $\xi$ is expected to be very small,
and weak pinning ($c\to0$), for which $\xi$ 
is directly related to the radial distribution function $g(r)$~\cite{jack-plaq-pin,jack-pin-chi4,szamel-pin2013}.

The key point is that
if the situation in Fig.~\ref{fig:dom}(a-c) holds in large systems,
bimodal distributions $p(Q)$ will be found on considering finite systems of size $L\lesssim\xi$.
The relevant distributions are sketched in Fig.~\ref{fig:dom}(d-f), and are similar to those in Figs.~\ref{fig:histo}(a,b).
Our results are therefore consistent with the inactive state having a correlation length $\xi\gtrsim L$. This situation occurs in plaquette
models at low temperatures~\cite{jack-plaq-pin,camma-patch}, where the spacing between localised `excitations'~\cite{GC-2003,GC-annrev-2010} 
determines the range of amorphous order.
It has also been proposed that a similar length scale determines 
the dynamical behaviour of non-equilibrium glass states formed by slow cooling~\cite{keys2013}.
Alternatively, the results of Fig.~\ref{fig:histo} are also consistent with the presence of a 
pinning-induced phase transition~\cite{biroli-pin,camma-pin-alpha,berthier-kob-pin2013}, in which case $\xi$ would be
infinite.  
In the absence of a finite-size scaling analysis, we cannot distinguish these two scenarios, so we simply conclude
that $\xi\gtrsim L$ for the inactive states that we consider.

To reinforce the connection between large domains and the results of Fig.~\ref{fig:histo}, 
it is useful to recall the implications of a non-convex
effective potential $V(Q,c)$, which necessarily accompanies any bimodal distribution $p(Q|c)$.  
The potential is non-convex if, for some $Q$, 
$(\partial^2/\partial Q^2)V(Q,c)<0$.  Hence, there exist two values of the overlap
$Q_1,Q_2$  such that $V(Q_1) + V(Q_2)  < 2V(Q_{\rm ave})$, where $Q_{\rm ave}=\frac12(Q_1+Q_2)$. Therefore
$p(Q_1) p(Q_2) > p(Q_{\rm ave})^2$. 
This means that systems which are globally high- or low-$q$ are more likely than
systems where the domains are mixed, which implies that 
domain sizes $\xi$ are comparable with the system size $L$~\cite{camma-phase-sep}.  Note that this argument holds for finite systems, independently of 
the existence of any phase transition: it is the behaviour of $V(Q)$ as $N\to\infty$ that determines phase behavior~\cite{berthier-silvio}.

Finally, we note that fluctuations of the overlap in these systems come from several sources: the choice of the
reference configuration and of which particles to pin (the fixed `template'), and the thermal fluctuations associated with the configuration 
$\CC_t$.  We find that the observed behaviour
differs significantly between different realisations of the template:
some templates are more likely to contribute to the large-$Q$ peaks
in Fig.~\ref{fig:histo} while other templates contribute more to the small-$Q$ peak.  In the picture of
Fig.~\ref{fig:dom}(a-c), this implies
that the high- or low-$q$ regions of space are tied to specific locations in the system, depending on the structure of the
template.
However, on varying the choice of the frozen particles for a given reference configuration, we do not find any strong
propensity for large-$Q$ or small-$Q$.  That is, the specific reference configuration does not strongly influence  
where the large-$q$ or small-$q$ domains in Fig.~\ref{fig:dom} are located.

To conclude, the results presented here indicate that inactive non-equilibrium states from the $s$-ensemble
have strong amorphous order, of a range $\xi$ comparable with the system size $L=5\sigma$.  This order is much stronger
that that found in equilibrium systems at the same temperature, consistent with the stability of the inactive states. 
The evidence
for the large length scale $\xi$ is indirect, but Fig.~\ref{fig:dom} shows how bimodal overlap distributions can be attributed
to the existence of large domains.
More generally, these results show how dynamical (non-equilibrium) $s$-ensembles~\cite{kcm-transition,hedges09}
can be combined with static concepts such as effective potentials~\cite{franz-parisi-1997} and amorphous 
order~\cite{BB-ktw-2004,montanari06,KL}, in order to understand stable glassy materials.

We thank Juan Garrahan, David Chandler, Ludovic Berthier
and Giulio Biroli for valuable discussions.  This work was funded by the EPSRC
through grant EP/I003797/1.


\begin{appendix}

\section{Supplementary Information for ``Investigating amorphous order in stable glasses by random pinning''}

\subsection{Model}

In the Kob-Andersen mixture, particles of types $\alpha,\beta$ 
interact by a Lennard-Jones potential $V_{\alpha\beta}(r) = 
4\epsilon_{\alpha\beta}[(\sigma_{\alpha\beta}/r)^12 - (\sigma_{\alpha\beta}/r)^6]$, which is
truncated and shifted at $r=2.5\sigma_{\alpha\beta}$ for numerical convenience.  The particle
types are labelled A and B and the interaction
parameters are $(\sigma_{\rm AA},\sigma_{\rm AB},\sigma_{\rm BB})=(1.0,0.80,0.88)\sigma$
and $(\epsilon_{\rm AA},\epsilon_{\rm AB},\epsilon_{\rm BB})=(1.0,1.5,0.5)\epsilon$.  
We consider $N=150$ particles with $N_{\rm A} = (4N/5)$ and $N_{\rm B} = (N/5)$. The system
is simulated by Monte Carlo dynamics~\cite{berthier-mc2007} with a maximal step in each
Cartesian direction of $0.075\sigma$, so the mean squared displacement for a single proposed move
is $(0.075\sigma)^2$.  Setting the time unit $\Delta t$ so that diffusion constant of a free particle is $D_0 = \sigma^2/\Delta t$,
one has the mean square displacement $\langle |\bm{r}(t)-\bm{r}(0)|^2 \rangle = 6 D_0 t = 6\sigma^2 (t/\Delta t)$.
Letting $t$ be the time interval associated with a single MC sweep (one attempted move per particle), one sees that 
the time $\Delta t$ corresponds to $6/(0.075)^2\approx 1070$ MC sweeps. 

\subsection{Inactive configurations}

The inactive configurations used in this work were obtained from an $s$-ensemble
constructed as in Refs.~\cite{hedges09, fullerton-veff}.  
A trajectory $\bm{r}^N(t)$ of length $t_{\rm obs}=M\Delta t$, has activity 
$K[\bm{r}^N(t)] = \Delta t\sum_{i=1}^{N_{\rm A}}\sum_{j=0}^{M}|\bm{r}_i(t_j) - \bm{r}_i(t_{j-1})|^2$,
where $t_j=j\Delta t$.  The corresponding ``intensive'' activity density is $k = K/(Nt_{\rm obs})$.
A biased ensemble of trajectories $\bm{r}^N(t)$ is defined through $P_s[\bm{r}^N(t)] =P_0[\bm{r}^N(t)]\exp\{-sK[\bm{r}^N(t)]\}/{\cal Z}(s)$,
where $s$ is a biasing field whose natural units are $(\sigma^2\Delta t)^{-1}$ and ${\cal Z}(s)$ is a normalisation constant.
We use transition path sampling~\cite{TPS}
to sample these $s$-ensembles, as in~\cite{hedges09,fullerton-veff}.

In particular, we generated trajectories of length $t_{\rm obs}=400\Delta t$ with
$s = 0.015$ [in units of $(\sigma^2\Delta t)^{-1}$], as in~\cite{fullerton-veff}.
The chosen value of $s$ corresponds to coexistence between the active and inactive
phases, allowing efficient sampling of trajectories from the inactive phase.
The inactive reference configurations that we use in this work were taken from 
trajectories in this $s$-ensemble: since the ensemble includes both dynamical phases, we take trajectories from the lower
third of the activity distribution as being typical of the inactive phase.
From these trajectories, the configurations at time $t_{\rm obs}/2$ form the set of configurations used as references: all results shown
involve averages over a sample
of $176$ independent configurations chosen in this way.  Several templates were generated from each of these configurations,
by independently choosing different sets of pinned particles.

\subsection{Ensembles with pinned particles}

Here, we give precise definitions of the ensembles that are associated with randomly pinned systems.
This situation has been analysed in detail by Krakoviack~\cite{krakoviack2010} but it is useful to 
review some results for the purposes of this work.  Our notation here follows similar work
for a spin model~\cite{jack-plaq-pin}.

Given a reference configuration $\CC$ with particle positions $\bm{r}_i$, 
we define a binary variable $f_i=0,1$ for each particle, where $f_i=1$
means that particle $i$ is pinned, and $f_i=0$ means that it is free to move.  Each $f_i$ is chosen
independently, having the value $1$ with probability $c$ and $0$ with probability $1-c$.
 
Together, the reference configuration $\CC$ and the variables $f_i$ encode the template, as described in the main text. 
Then, consider an ensemble of configurations $\CC'$ that are consistent with the template, with weights
according to the Boltzmann distribution.  That is, if the particle positions in $\CC'$ are $\bm{r}_i'$ then
\begin{equation}
P(\CC'|\CC,\{f_i\}) =\frac{1}{Z(\CC,\{f_i\})} \ee^{-\beta E(\CC')} \prod_{j\in \cal F} \delta( \bm{r}_j' - \bm{r}_j ) 
\label{equ:dist-cc}
\end{equation}
where $E(\CC')$ is the energy of configuration $\CC'$, while 
$Z(\CC,\{f_i\})=\int\mathrm{d}\bm{r}''^N \ee^{-\beta E(\CC'')} \prod_{j\in \cal F} \delta( \bm{r}_j'' - \bm{r}_j )$ 
is a normalisation constant (partition function), and $\cal F$ is the set of particles $j$ 
with $f_j=1$.  In some situations (for example a perturbative analysis at small-$c$~\cite{jack-plaq-pin}), it may be useful to
write $\prod_{j\in \cal F} \delta( \bm{r}_j' - \bm{r}_j ) = \prod_j [ (1-f_j) + f_j \delta( \bm{r}_j' - \bm{r}_j ) ]$
where the product on the right hand side now runs over all particles.  This equality holds because the only possible
values of $f_j$ are zero and unity.

In the dynamical simulations presented here, the Monte Carlo algorithm respects detailed balance
with respect to the distribution (\ref{equ:dist-cc}), so on taking $t\to\infty$ for a given template,
the configurations $\CC_t$ generated by dynamical simulation must converge to 
the distribution of (\ref{equ:dist-cc}) with $\CC'=\CC_t$.  
By sampling templates from a given distribution (equilibrium or inactive) and taking $t\to\infty$,
one may therefore sample the joint distribution of $\CC'$ with the template. 
These joint distributions $P(\CC',\CC,\{f_i\})$ can be used to calculate the results of the main text: in particular,
$p_{\rm eq}(Q|c)$ is the marginal distribution of $Q(\CC,\CC')$ obtained from a joint distribution $P(\CC',\CC,\{f_i\})$
that is formed by using (\ref{equ:dist-cc}) in conjunction with an equilibrated
distribution for $\CC$, and with $f_i$ chosen independently as described above.  Similarly, $p_{\rm in}(Q|c)$
is a similar marginal but with $P(\CC',\CC,\{f_i\})$ constructed by drawing $\CC$ from the inactive state.

For example, if $\CC$ is chosen from an equilibrium state then the joint distribution of the template
and the configuration $\CC'$ is
\begin{multline}
P_{\rm eq}(\CC',\CC,\{f_i\}) = \frac{1}{Z_f} \ee^{-\mu_{\rm f}\sum_i f_i} \\ \times \frac{1}{Z_2(\{f_i\})}
  \ee^{-\beta E(\CC)-\beta E(\CC')} \prod_{j\in \cal F} \delta( \bm{r}_j' - \bm{r}_j ) 
  \label{equ:pcc'}
\end{multline}
where $Z_f = (1+\ee^{-\mu_{\rm f}})^N$ and $Z_2(\{f_i\}) = \int\mathrm{d}\bm{r}^N \mathrm{d}\bm{r}'^N
\ee^{-\beta E(\CC)-\beta E(\CC')} \prod_{j\in \cal F} \delta( \bm{r}_j - \bm{r}'_j ) $.

For equilibrium pinning, we note that (\ref{equ:pcc'}) is symmetric in $\CC$ and $\CC'$, and the marginal distribution of $\CC$ is
the equilibrium distribution, by construction.  Hence the distribution of $\CC'$ is also the equilibrium
distribution of the system: that is, the structure of $\CC'$ is unaffected by the pinning.
On the other hand, if $\CC$ does not have an equilibrium distribution, as for the inactive reference states
considered here, then the distribution of $\CC'$ is not equilibrated, nor is it equal to the distribution
of $\CC$.  Rather, it represents a system that has biased away from equilibrium and 
towards to the inactive state, through the presence of the template.


\end{appendix}

\end{document}